\documentclass{IEEEtran}
\makeatletter

\ifCLASSOPTIONcompsoc
  \usepackage[caption=false,font=normalsize,labelfont=sf,textfont=sf]{subfig}
\else
  \usepackage[caption=false,font=footnotesize]{subfig}
\fi

\usepackage[techreport]{optional}

\usepackage{fmtcount} 
\usepackage{graphicx, amsmath, amssymb, subfig}
\usepackage{algorithm}
\usepackage{algorithmic} 
\usepackage{url,times}

\newtheorem{proposition}{\bf Proposition}

\newtheorem{lemma}{\bf Lemma}
\newtheorem{definition}{\bf Definition}

\DeclareMathOperator*{\argmax}{arg\,max}

\newcommand{\vD}{\mathbf{D}}
\newcommand{\vA}{\mathbf{A}}
\newcommand{\vd}{\mathbf{d}}
\newcommand{\vc}{\mathbf{c}}

\newcommand{\vx}{\mathbf{x}}
\newcommand{\vy}{\mathbf{y}}
\newcommand\norm[1]{\left\lVert#1\right\rVert}

\def\squareforqed{\IEEEQED}
\def\qed{\ifmmode\squareforqed\else{\unskip\nobreak\hfil
\penalty50\hskip1em\null\nobreak\hfil\squareforqed
\parfillskip=0pt\finalhyphendemerits=0\endgraf}\fi}

\def\hollowsquare{\hbox{\rlap{$\sqcap$}$\sqcup$}}
\def\hollowqed{\ifmmode\hollowsquare\else{\unskip\nobreak\hfil
\penalty50\hskip1em\null\nobreak\hfil\hollowsquare
\parfillskip=0pt\finalhyphendemerits=0\endgraf}\fi}

\begin{document}

\title{Dominant Resource Fairness in Cloud Computing Systems with Heterogeneous Servers}
  
\author{
\IEEEauthorblockN{Wei Wang, Baochun Li, Ben Liang\\}
\IEEEauthorblockA{Department of Electrical and Computer Engineering\\
University of Toronto}
\vspace*{-0.3in}
}
	
\maketitle

\begin{abstract}

We study the multi-resource allocation problem in cloud computing systems where the resource pool is constructed from a large number of heterogeneous servers, representing different points in the configuration space of resources such as processing, memory, and storage. We design a multi-resource allocation mechanism, called DRFH, that generalizes the notion of Dominant Resource Fairness (DRF) from a single server to multiple heterogeneous servers. DRFH provides a number of highly desirable properties. With DRFH, no user prefers the allocation of another user; no one can improve its allocation without decreasing that of the others; and more importantly, no user has an incentive to lie about its resource demand. As a direct application, we design a simple heuristic that implements DRFH in real-world systems. Large-scale simulations driven by Google cluster traces show that DRFH significantly outperforms the traditional slot-based scheduler, leading to much higher resource utilization with substantially shorter job completion times.

\end{abstract}

\section{Introduction}
\label{sec:intro}


Resource allocation under the notion of fairness and efficiency is a fundamental problem in the design of cloud computing systems. Unlike traditional application-specific clusters and grids, a cloud computing system distinguishes itself with unprecedented server and workload heterogeneity. Modern datacenters are likely to be constructed from a variety of server classes, with different configurations in terms of processing capabilities, memory sizes, and storage spaces \cite{Armbrust10a}. Asynchronous hardware upgrades, such as adding new servers and phasing out existing ones, further aggravate such diversity, leading to a wide range of server specifications in a cloud computing system \cite{Reiss12a}. Table~\ref{tbl:server-config} illustrates the heterogeneity of servers in one of Google's clusters \cite{Reiss12a, GoogleTrace}.

In addition to server heterogeneity, cloud computing systems also represent much higher diversity in resource demand profiles. Depending on the underlying applications, the workload spanning multiple cloud users may require vastly different amounts of resources ({\em e.g.}, CPU, memory, and storage). For example, numerical computing tasks are usually CPU intensive, while database operations typically require high-memory support. The heterogeneity of both servers and workload demands poses significant technical challenges on the resource allocation mechanism, giving rise to many delicate issues --- notably fairness and efficiency --- that must be carefully addressed.

Despite the unprecedented heterogeneity in cloud computing systems,
state-of-the-art computing frameworks employ rather simple abstractions that
fall short. For example, Hadoop \cite{Hadoop} and Dryad \cite{Isard07a}, the
two most widely deployed cloud computing frameworks, partition a server's
resources into bundles --- known as {\em slots} --- that contain fixed amounts
of different resources. The system then allocates resources to users at the
granularity of these slots. Such a {\em single resource} abstraction ignores
the heterogeneity of both server specifications and demand profiles, inevitably
leading to a fairly inefficient allocation \cite{Ghodsi11a}.

\begin{table}[t]
   \centering
   \renewcommand{\arraystretch}{0.99}
   \footnotesize
   \caption{Configurations of servers in one of Google's clusters
	   \cite{Reiss12a, GoogleTrace}. CPU and memory units are normalized to
	   the maximum server (highlighted below).}
   \vspace{-2mm}
   \begin{tabular}{|c||c|c|}
     \hline
     {\bf Number of servers} & {\bf CPUs} & {\bf Memory} \\
     \hline
     6732 & 0.50 & 0.50 \\
     \hline
     3863 & 0.50 & 0.25 \\
     \hline
     1001 & 0.50 & 0.75 \\
     \hline
     795 & {\bf 1.00} & {\bf 1.00} \\
     \hline
     126 & 0.25 & 0.25 \\
     \hline
     52 & 0.50 & 0.12 \\
     \hline
     5 & 0.50 & 0.03 \\
     \hline
     5 & 0.50 & 0.97 \\
     \hline
     3 & 1.00 & 0.50 \\
     \hline
     1 & 0.50 & 0.06 \\
     \hline
   \end{tabular}
   \label{tbl:server-config}
   \vspace{-.1in}
\end{table}

Towards addressing the inefficiency of the current allocation system, many recent works focus on {\em multi-resource allocation} mechanisms. Notably, Ghodsi {\em et al.}~\cite{Ghodsi11a} suggest a compelling alternative known as the Dominant Resource Fairness (DRF) allocation, in which each user's {\em dominant share} --- the maximum ratio of any resource that the user has been allocated in a server --- is equalized. The DRF allocation possesses a set of highly desirable fairness properties, and has quickly received significant attention in the literature \cite{Joe-Wong12a, Dolev12a, Gutman12a, Parkes12a}. While DRF and its subsequent works address the demand heterogeneity of multiple resources, they all ignore the heterogeneity of servers, limiting the discussions to a hypothetical scenario where all resources are concentrated in one super computer\footnote{While \cite{Ghodsi11a} briefly touches on the case where resources are distributed to small servers (known as the discrete scenario), its coverage is rather informal.}. Such an {\em all-in-one} resource model drastically contrasts the state-of-the-practice infrastructure of cloud computing systems. In fact, with heterogeneous servers, even the definition of {\em dominant resource} is unclear: Depending on the underlying server configurations, a computing task may bottleneck on different resources in different servers. We shall note that naive extensions, such as applying the DRF allocation to each server separately, leads to a highly inefficient allocation (details in Sec.~\ref{sec:drf}).

This paper represents the first rigorous study to propose a solution with
provable operational benefits that bridge the gap between the existing
multi-resource allocation models and the prevalent datacenter infrastructure.
We propose {\bf DRFH}, a generalization of {\bf DRF} mechanism in {\bf
	H}eterogeneous environments where resources are pooled by a large
amount of heterogeneous servers, representing different points in the
configuration space of resources such as processing, memory, and storage. DRFH generalizes the
intuition of DRF by seeking an allocation that equalizes every user's {\em
	global dominant share}, which is the maximum ratio of any resources the
user has been allocated in the {\em entire} cloud resource pool. We
systematically analyze DRFH and show that it retains most of the desirable
properties that the all-in-one DRF model provides \cite{Ghodsi11a}.
Specifically, DRFH is {\em Pareto optimal}, where no user is able to increase
its allocation without decreasing other users' allocations. Meanwhile, DRFH is
{\em envy-free} in that no user prefers the allocation of another user. More
importantly, DRFH is {\em truthful} in that a user cannot schedule more
computing tasks by claiming more resources that are not needed, and hence has
no incentive to misreport its actual resource demand. DRFH also satisfies a set
of other important properties, namely {\em single-server DRF}, {\em
	single-resource fairness}, {\em bottleneck fairness}, and {\em
	population monotonicity} (details in Sec.~\ref{sec:properties}).

As a direct application, we design a heuristic scheduling algorithm that
implements DRFH in real-world systems. We conduct large-scale simulations
driven by Google cluster traces \cite{GoogleTrace}. Our simulation results show
that compared to the traditional slot schedulers adopted in prevalent cloud
computing frameworks, the DRFH algorithm suitably matches demand heterogeneity
to server heterogeneity, significantly improving the system's resource
utilization, yet with a substantial reduction of job completion times.


\section{Related Work}
\label{sec:related}

Despite the extensive computing system literature on fair resource allocation,
much of the existing works limit their discussions to the allocation of a
single resource type, {\em e.g.}, CPU time \cite{Baruah95a, Baruah96a} and link
bandwidth \cite{Kelly98a, Mo00a, Kleinberg99a, Blanquer01a, Liu03a}. Various
fairness notions have also been proposed throughout the years, ranging from
application-specific allocations \cite{Koksal00a, Bredel09a} to general
fairness measures \cite{Kelly98a, Jain84a, Lan10a}.

As for multi-resource allocation, state-of-the-art cloud computing systems
employ naive single resource abstractions. For example, the two fair sharing
schedulers currently supported in Hadoop \cite{HadoopCS, HadoopFS} partition a
node into slots with fixed fractions of resources, and allocate resources
jointly at the slot granularity. Quincy \cite{Isard09a}, a fair scheduler
developed for Dryad \cite{Isard07a}, models the fair scheduling problem as a
min-cost flow problem to schedule jobs into slots. The recent work
\cite{Ghodsi13a} takes the job placement constraints into consideration, yet 
it still uses a slot-based single resource abstraction.

Ghodsi {\em et al.}~\cite{Ghodsi11a} are the first in the literature to present
a systematic investigation on the multi-resource allocation problem in cloud
computing systems. They propose DRF to equalize the dominant share of all
users, and show that a number of desirable fairness properties are guaranteed
in the resulting allocation. 
DRF has quickly attracted a substantial amount of attention and has been
generalized to many dimensions. Notably, Joe-Wong {\em et
al.}~\cite{Joe-Wong12a} generalize the DRF measure and incorporate it into a
unifying framework that captures the trade-offs between allocation fairness and
efficiency. Dolev {\em et al.}~\cite{Dolev12a} suggest another notion of
fairness for multi-resource allocation, known as Bottleneck-Based Fairness
(BBF), under which two fairness properties that DRF possesses are also
guaranteed. Gutman and Nisan \cite{Gutman12a} consider another settings of DRF
with a more general domain of user utilities, and show their connections to the
BBF mechanism. Parkes {\em et al.}~\cite{Parkes12a}, on the other hand, extend
DRF in several ways, including the presence of zero demands for certain
resources, weighted user endowments, and in particular the case of indivisible
tasks. They also study the loss of social welfare under the DRF rules. More
recently, the ongoing work of Kash {\em et al.}~\cite{Kash12a} extends the DRF
model to a dynamic setting where users may join the system over time but will
never leave. Though motivated by the resource allocation problem in cloud
computing systems, all the works above restrict their discussions to a
hypothetical scenario where the resource pool contains only one big server,
which is not the case in the state-of-the-practice datacenter systems.

Other related works include fair-division problems in the economics literature, in particular the {\em egalitarian division} under Leontief preferences \cite{Li11a} and the {\em cake-cutting problem} \cite{Procaccia13a}. However, these works also assume the {\em all-in-one} resource model, and hence cannot be directly applied to cloud computing systems with heterogeneous servers. 

\section{System Model and Allocation Properties}
\label{sec:model}

In this section, we model multi-resource allocation in a cloud computing system
with heterogeneous servers. We formalize a number of desirable properties that
are deemed the most important for allocation mechanisms in cloud computing
environments.

\subsection{Basic Setting}
\label{sec:setting}

In a cloud computing system, the resource pool is composed of a cluster
of heterogeneous servers $S = \{ 1, \dots, k \}$, each contributing $m$ hardware
resources ({\em e.g.}, CPU, memory, storage) denoted by $R = \{ 1,\dots,m \}$. 
For each server $l$, let $\vc_l = ( c_{l1},
\dots, c_{lm})^T$ be its {\em resource capacity vector}, where each
element $c_{lr}$ denotes the total amount of resource $r$ available in server $l$.
Without loss of generality, for every resource $r$, we normalize the total
capacity of all servers to 1, {\em i.e.}, 
\begin{equation*}
  \sum_{l \in S} c_{lr} = 1, \quad r = 1, 2, \ldots, m.
\end{equation*}

Let $U = \{ 1,\dots,n \}$ be the set of cloud users sharing the cloud
system. For every user $i$, let $\vD_i = ( D_{i1}, \dots, D_{im})^T$ be its
{\em resource demand vector}, where $D_{ir}$ is the fraction (share) of resource $r$
required by each task of user $i$ over the {\em entire} system. For simplicity, we assume positive demands for
all users, {\em i.e.}, $D_{ir} > 0, \forall i \in U, r \in R$. We say resource
$r^*_i$ is the {\em global dominant resource} of user $i$ if
\begin{equation*}
  r^*_i \in \argmax_{r \in R} D_{ir}~.
\end{equation*}
In other words, $r^*_i$ is the most heavily demanded resource required by 
user $i$'s task in the entire resource pool. For all user $i$ and resource $r$,
we define
\begin{equation*}
	d_{ir} = D_{ir} / D_{ir^*_i}
\end{equation*}
as the {\em normalized demand} and denote by $\vd_i = ( d_{i1}, \dots,
d_{im})^T$ the normalized demand vector of user $i$. 

\begin{figure}[tb]
  \centering
  \includegraphics[width=0.3\textwidth]{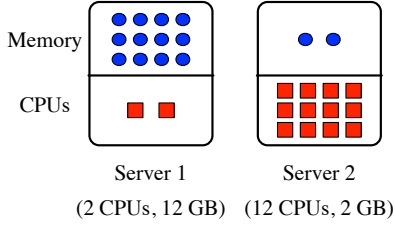}
  \vspace{-.1in}
  \caption{An example of a system containing two heterogeneous
  servers shared by two users. Each computing task of user 1 requires 0.2 CPU time and 1 GB memory,
  while the computing task of user 2 requires 1 CPU time and 0.2 GB memory.
  }
  \label{fig:drf-not-po}
  \vspace{-.1in}
\end{figure}

As a concrete example, consider Fig.~\ref{fig:drf-not-po} where the 
system contains two heterogeneous servers. Server 1 is high-memory with 2
CPUs and 12 GB memory, while server 2 is high-CPU with 12 CPUs and 2 GB
memory.  Since the system contains 14 CPUs and 14 GB memory in total, the
normalized capacity vectors of server 1 and 2 are $\vc_1 = (\mbox{CPU share},
\mbox{memory share})^T = (1/7, 6/7)^T$ and $\vc_2 = (6/7, 1/7)^T$,
respectively.  Now suppose there are two users. User 1 has memory-intensive
tasks each requiring 0.2 CPU time and 1 GB memory, while user 2 has CPU-heavy
tasks each requiring 1 CPU time and 0.2 GB memory.  The demand vector of user 1
is $\vD_1 = (1/70, 1/14)^T$ and the normalized vector is $\vd_1 = (1/5, 1)^T$,
where memory is the global dominant resource. Similarly, user 2 has
$\vD_2 = (1/14, 1/70)^T$ and $\vd_2 = (1, 1/5)^T$, and CPU is its global dominant
resource.


For now, we assume users have an infinite number of tasks to be scheduled, and all
tasks are divisible \cite{Ghodsi11a, Dolev12a, Gutman12a, Parkes12a, Kash12a}.
We will discuss how these assumptions can be relaxed in Sec.~\ref{sec:prac}.

\subsection{Resource Allocation}
\label{sec:res-alloc}

For every user $i$ and server $l$, let $\vA_{il} = ( A_{il1}, \dots, A_{ilm}
)^T$ be the {\em resource allocation vector}, where $A_{ilr}$ is the share of
resource $r$ allocated to user $i$ in server $l$. Let $\vA_i = ( \vA_{i1},
\dots, \vA_{ik} )$ be the {\em allocation matrix} of user $i$, and $\vA = (
\vA_1, \dots, \vA_n )$ the overall allocation for all users. We say an
allocation $\vA$ is {\em feasible} if no server is required to use more than any of
its total resources, {\em i.e.},
\begin{equation*}
  \sum_{i \in U} A_{ilr} \le c_{lr} , \quad \forall l \in S, r \in R~. 
\end{equation*}

For all user $i$, given allocation $\vA_{il}$ in server $l$, the maximum
number of tasks (possibly fractional) that it can schedule is calculated as
\begin{equation*}
	N_{il}(\vA_{il}) = \min_{r \in R} \{ A_{ilr} / D_{ir} \}~.
\end{equation*} 
The total number of tasks user $i$ can schedule under allocation $\vA_i$ is hence
\begin{equation}
  \label{eq:util-overl}
  N_i(\vA_i) = \sum_{l \in S}  N_{il}(\vA_{il})~.
\end{equation}
Intuitively, a user prefers an allocation that allows it to schedule more tasks.

A well-justified allocation should never give a user more resources than it can
actually use in a server. Following the terminology used in the economics
literature \cite{Li11a}, we call such an allocation {\em non-wasteful}:
\begin{definition}
  \label{def:nonwaste}
  For user $i$ and server $l$, an allocation $\vA_{il}$ is {\em non-wasteful}
  if taking out any resources reduces the number of tasks scheduled, {\em i.e.,}
  for all $\vA'_{il} \prec \vA_{il}$\footnote{For any two vectors $\vx$ and
	  $\vy$, we say $\vx \prec \vy$ if $x_i \le y_i, \forall i$ and for
	  some $j$ we have strict inequality: $x_j < y_j$.}, we have that
  \begin{equation*}
	  N_{il}(\vA'_{il}) < N_{il}(\vA_{il})~.
  \end{equation*}
  User $i$'s allocation $\vA_i
  = (\vA_{il})$ is non-wasteful if $\vA_{il}$ is non-wasteful for all
  server $l$, and allocation $\vA = (\vA_i)$ is non-wasteful if $\vA_i$
  is non-wasteful for all user $i$.
\end{definition}

Note that one can always convert an allocation to non-wasteful by revoking
those resources that are allocated but have never been actually used, without
changing the number of tasks scheduled for any user. Therefore, unless
otherwise specified, we limit the discussions to non-wasteful allocations.

\subsection{Allocation Mechanism and Desirable Properties}
\label{sec:properties}

A resource allocation mechanism takes user demands as input and outputs the
allocation result. In general, an allocation mechanism should provide the
following essential properties that are widely recognized as the most
important fairness and efficiency measures in both cloud computing systems
\cite{Ghodsi11a, Joe-Wong12a, Ghodsi13a} and the economics literature
\cite{Li11a, Procaccia13a}.

{\em Envy-freeness:} An allocation mechanism is {\em envy-free} if no user
prefers the other's allocation to its own, {\em i.e.}, $N_i(\vA_i) \ge N_i(\vA_j)$
for any two users $i,j \in U$. This property essentially embodies the notion of fairness.

{\em Pareto optimality:} An allocation mechanism is {\em Pareto optimal} if it
returns an allocation $\vA$ such that for all feasible allocations $\vA'$, if
$N_i(\vA'_i) > N_i(\vA_i)$ for some user $i$, then there exists a user $j$ such
that $N_j(\vA'_j) < N_j(\vA_j)$. In other words, there is no other allocation where
all users are at least as well off and at least one user is strictly better
off. This property ensures the allocation efficiency and is critical 
for high resource utilization.

{\em Truthfulness:} An allocation mechanism is {\em truthful} if no user can
schedule more tasks by misreporting its resource demand (assuming a user's demand is its private information), irrespective of other
users' behaviour.
Specifically, given the demands claimed by other users, let $\vA$ be the resulting
allocation when user $i$ truthfully reports its resource demand $\vD_i$, and
let $\vA'$ be the allocation returned when user $i$ misreports by $\vD'_i \neq
\vD_i$. Then under a truthful mechanism we have $N_i(\vA_i) \ge N_i(\vA'_i)$.
Truthfulness is of a special importance for a cloud computing system, as it is common to
observe in real-world systems that users try to lie about their resource demands to
manipulate the schedulers for more allocation \cite{Ghodsi11a, Ghodsi13a}.

In addition to these essential properties, we also consider four other
important properties below:

{\em Single-server DRF:} If the system contains only one server, then the
resulting allocation should be reduced to the DRF allocation.

{\em Single-resource fairness:} If there is a single resource in the system,
then the resulting allocation should be reduced to a max-min fair allocation.

{\em Bottleneck fairness:} If all users bottleneck on the same resource ({\em i.e.,} having
the same global dominant resource), then the resulting allocation should be
reduced to a max-min fair allocation for that resource.

{\em Population monotonicity:} If a user leaves the system and relinquishes all
its allocations, then the remaining users will not see any reduction in the
number of tasks scheduled.

In addition to the aforementioned properties, {\em sharing incentive} is
another important property that has been frequently mentioned in the literature \cite{Ghodsi11a,
	Joe-Wong12a, Dolev12a, Parkes12a}. It ensures that every user's
allocation is not worse off than that obtained by evenly dividing the entire
resource pool. While this property is well defined for a single server, it is
not for a system containing multiple heterogeneous servers, as there is an
infinite number of ways to evenly divide the resource pool among users, and it is unclear
which one should be chosen as a benchmark. We
defer the discussions to Sec.~\ref{sec:sharing-incentive}, where we justify between two
possible alternatives. For now, our objective is to design an allocation
mechanism that guarantees all the properties defined above. 



\subsection{Naive DRF Extension and Its Inefficiency}
\label{sec:drf}

\begin{figure}[tb]
  \centering
  \includegraphics[width=0.35\textwidth]{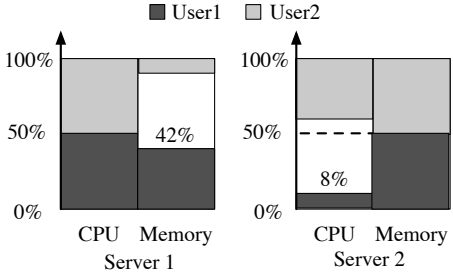}
  \vspace{-.1in}
  \caption{DRF allocation for the example shown in Fig.~\ref{fig:drf-not-po}, where
  user 1 is allocated 5 tasks in server 1 and 1 in server 2, while user 2 is
  allocated 1 task in server 1 and 5 in server 2.
  }
  \label{fig:drf-alloc}
  \vspace{-.1in}
\end{figure}

It has been shown in \cite{Ghodsi11a, Parkes12a} that the DRF allocation
satisfies all the desirable properties mentioned above when there is only one server
in the system. The key intuition is to equalize the fraction of dominant
resources allocated to each user in the server. When resources are distributed
to many heterogeneous servers, a naive generalization is to separately apply
the DRF allocation per server. Since servers are heterogeneous, a user might
have different dominant resources in different servers. For instance, in the example
of Fig.~\ref{fig:drf-not-po}, user 1's dominant resource in server 1 is
CPU, while its dominant resource in server 2 is memory. Now apply DRF in
server 1. Because CPU is also user 2's dominant resource, the DRF allocation
lets both users have an equal share of the server's CPUs, each allocated 1. As a
result, user 1 schedules 5 tasks onto server 1, while user 2 schedules 1 onto the
same server. Similarly, in server 2, memory is the dominant resource of both
users and is evenly allocated, leading to 1 task scheduled for user 1 and 5 for
user 2.  The resulting allocations in the two servers are illustrated in
Fig.~\ref{fig:drf-alloc}, where both users schedule 6 tasks. 

Unfortunately, this allocation violates Pareto optimality and is highly
inefficient. If we instead allocate server 1 exclusively to user 1, and
server 2 exclusively to user 2, then both users schedule 10 tasks, more than
those scheduled under the DRF allocation. In fact, we see that naively applying
DRF per server may lead to an allocation with arbitrarily low resource
utilization.

The failure of the naive DRF extension to the heterogeneous environment
necessitates an alternative allocation mechanism, which is the main theme of the next section.

\section{DRFH Allocation and Its Properties}
\label{sec:drfc}

In this section, we describe DRFH, a generalization of DRF in a heterogeneous
cloud computing system where resources are distributed in a number of
heterogeneous servers. We analyze DRFH and show that it provides all the
desirable properties defined in Sec.~\ref{sec:model}.

\subsection{DRFH Allocation}

Instead of allocating separately in each server, DRFH jointly considers
resource allocation across all heterogeneous servers. The key intuition is to
achieve the {\em max-min fair allocation for the global dominant resources}.
Specifically, given allocation $\vA_{il}$, let
$G_{il}(\vA_{il})$ be the fraction of global dominant resources user $i$
receives in server $l$, {\em i.e.,} 
\begin{equation}
	\label{eq:G_i_l}
	G_{il}(\vA_{il}) = N_{il}(\vA_{il}) D_{i r_i^*} = \min_{r \in R} \{ A_{ilr} / d_{ir} \}~.
\end{equation}
We call $G_{il}(\vA_{il})$ the {\em global dominant share} user $i$ receives in
server $l$ under allocation $\vA_{il}$. Therefore, given the overall allocation $\vA_i$, the global dominant share
user $i$ receives is 
\begin{equation}
	\label{eq:G_i}
	G_i(\vA_i) = \sum_{l \in S} G_{il}(\vA_{il}) = \sum_{l \in S} \min_{r \in R} \{ A_{ilr} / d_{ir} \}~.
\end{equation}
DRFH allocation aims to maximize the minimum global dominant
share among all users, subject to the resource constraints per server, {\em i.e.,}
\begin{equation}
  \label{eq:max-min}
  \begin{split}
	  \max_{\vA} & \quad \min_{i \in U} G_i(\vA_i) \\
    \mbox{s.t.} & \quad \sum_{i \in U} A_{ilr} \le c_{lr}, \forall l \in S, r \in R~.
  \end{split}
\end{equation}

Recall that without loss of generality, we assume non-wasteful allocation $\vA$
(see Sec.~\ref{sec:res-alloc}). We have the following structural
result. \opt{conf}{The proof is deferred to our technical report \cite{Wang-DRFH}.}

\begin{lemma}
  \label{lem:nonwaste}
  For user $i$ and server $l$, an allocation $\vA_{il}$ is non-wasteful {\em
	  if and only if} there exists some $g_{il}$ such that $\vA_{il} =
  g_{il} \vd_i$. In particular, $g_{il}$ is the global dominant share user $i$
  receives in server $l$ under allocation $\vA_{il}$, {\em i.e.,}
  \begin{equation*}
	  g_{il} = G_{il}(\vA_{il})~.
  \end{equation*}
\end{lemma}

\opt{techreport}{
	{\bf Proof:}
	($\Leftarrow$) We start with the necessity proof.
	Since $\vA_{il} = g_{il} \vd_i$, for all resource $r \in R$, we have
	\begin{equation*}
		A_{ilr} / D_{ir} = g_{il} d_{ir} / D_{ir} = g_{il} D_{i r_i^*}~.
	\end{equation*}
	As a result, 
	\begin{equation*}
		N_{il}(\vA_{il}) = \min_{r \in R} \{ A_{ilr} / D_{ir} \} 
		= g_{il} D_{i r_i^*}~.
	\end{equation*}
	Now for any $\vA'_{il} \prec \vA_{il}$, suppose $A'_{ilr_0} < A_{ilr_0}$ for some resource
	$r_0$. We have
	\begin{align*}
		N_{il}(\vA'_{il}) & = \min_{r \in R} \{ A'_{ilr} / D_{ir} \} \\
		& \le A'_{ilr_0} / D_{i r_0} \\
		& < A_{ilr_0} / D_{i r_0} = N_{il}(\vA_{il})~.
	\end{align*}
	Hence by definition, allocation $\vA_{il}$ is non-wasteful.

	($\Rightarrow$) We next present the sufficiency proof. Since $\vA_{il}$ is non-wasteful,
	for any two resources $r_1, r_2 \in R$, we must have
	\begin{equation*}
		A_{il r_1} / D_{i r_1} = A_{il r_2} / D_{i r_2}~.
	\end{equation*}
	Otherwise, without loss of generality, suppose $A_{il r_1} / D_{i r_1} > A_{il r_2} / D_{i r_2}$.
	There must exist some $\epsilon > 0$, such that
	\begin{equation*}
		(A_{il r_1} - \epsilon) / D_{i r_1} > A_{il r_2} / D_{i r_2}~.
	\end{equation*}
	Now construct an allocation $\vA'_{il}$, such that
	\begin{equation}
		A'_{ilr} = \left\{
			\begin{array}{cc}
				A'_{il r_1} - \epsilon, & \quad r = r_1; \\
				A_{ilr}, & \mbox{o.w.}
			\end{array}
			\right.
	\end{equation}
	Clearly, $\vA'_{il} \prec \vA_{il}$. However,
	it is easy to see that
	\begin{align*}
		N_{il}(\vA'_{il}) & = \min_{r \in R} \{ A'_{ilr} / D_{ir} \} \\
		& = \min_{r \neq r_1} \{ A'_{ilr} / D_{ir} \} \\
		& = \min_{r \neq r_1} \{ A_{ilr} / D_{ir} \} \\
		& = \min_{r \in R} \{ A_{ilr} / D_{ir} \} = N_{il}(\vA_{il}) ~,
	\end{align*}
	which contradicts the fact that $\vA_{il}$ is non-wasteful.
	As a result, there exits some $n_{il}$, such that for all resource $r \in R$, we have
	\begin{equation*}
		A_{ilr}= n_{il} D_{ir} = n_{il} D_{i r_i^*} d_{ir}~.
	\end{equation*}
	Now letting $g_{il} = n_{il} D_{i r_i^*}$, we see $\vA_{il} = g_{il} \vd_i$.
	\qed
}

Intuitively, Lemma~\ref{lem:nonwaste} indicates that under a non-wasteful
allocation, resources are allocated {\em in proportion} to the user's
demand.  Lemma~\ref{lem:nonwaste} immediately suggests the following
relationship for all user $i$ and its non-wasteful allocation $\vA_i$:
\begin{equation*}
  G_i(\vA_i) = \sum_{l \in S} G_{il}(\vA_{il}) = 
  \sum_{l \in S} g_{il}~.
\end{equation*}
Problem \eqref{eq:max-min} can hence be equivalently written as
\begin{equation}
  \label{eq:maxmin-fair}
  \begin{split}
	  \max_{\{g_{il}\}} & \quad \min_{i \in U} \sum_{l \in S} g_{il} \\
    \mbox{s.t.} & \quad \sum_{i \in U} g_{il} d_{ir} \le c_{lr}, \forall l \in S, r \in R~,
  \end{split}
\end{equation}
where the constraints are derived from Lemma~\ref{lem:nonwaste}. 
Now let $g = \min_{i} \sum_{l \in S} g_{il}$. Via straightforward algebraic operation,
we see that (\ref{eq:maxmin-fair}) is equivalent to the following problem:
\begin{equation}
  \label{eq:h-drf}
  \begin{split}
	  \max_{\{g_{il}\}} & \quad g \\
    \mbox{s.t.} & \quad \sum_{i \in U} g_{il} d_{ir} \le c_{lr}, \forall l \in S, r \in R~, \\
    & \quad \sum_{l \in U} g_{il} = g, \forall i \in U~.
  \end{split}
\end{equation}
Note that the second constraint ensures the fairness with respect to the
equalized global dominant share $g$. By solving (\ref{eq:h-drf}), DRFH
allocates each user the maximum global dominant share $g$, under the
constraints of both server capacity and fairness. By Lemma~\ref{lem:nonwaste},
the allocation received by each user $i$ in server $l$ is simply $\vA_{il} = g_{il} \vd_i$.

\begin{figure}[tb]
  \centering
  \includegraphics[width=0.37\textwidth]{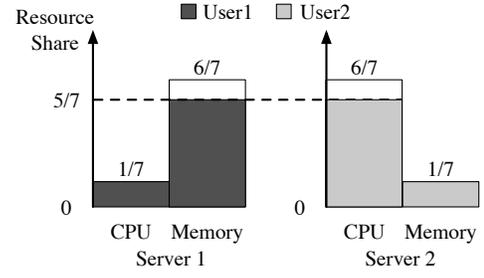}
  \vspace{-.1in}
  \caption{An alternative allocation with higher system utilization for the
  example of Fig.~\ref{fig:drf-not-po}. Server 1 and 2 are exclusively
  assigned to user 1 and 2, respectively. Both users schedule 10 tasks.
  }
  \label{fig:h-drf-alloc}
  \vspace{-.1in}
\end{figure}

For example, Fig.~\ref{fig:h-drf-alloc} illustrates the resulting DRFH
allocation in the example of Fig.~\ref{fig:drf-not-po}. By solving
(\ref{eq:h-drf}), DRFH allocates server 1 exclusively to user 1 and server 2
exclusively to user 2, allowing each user to schedule 10 tasks with the maximum
global dominant share $g = 5/7$.

We next analyze the properties of DRFH allocation obtained by solving
(\ref{eq:h-drf}) in the following two subsections. 

\subsection{Analysis of Essential Properties}

Our analysis of DRFH starts with the three essential resource allocation
properties, namely, envy-freeness, Pareto optimality, and truthfulness. We
first show that under the DRFH allocation, no user prefers other's allocation
to its own.

\begin{proposition}[Envy-freeness]
  The DRFH allocation obtained by solving (\ref{eq:h-drf}) is envy-free.
\end{proposition}

{\bf Proof:}
  Let $\{ g_{il} \}$ be the solution to problem (\ref{eq:h-drf}). For all
  user $i$, its DRFH allocation in server $l$ is $\vA_{il} = g_{il} \vd_i$. To show
  $N_i(\vA_j) \le N_i(\vA_i)$ for any two users $i$ and $j$, it is equivalent
  to prove $N_i(\vA_j) \le N_i(\vA_i)$. We have
  \begin{align*}
      G_i(\vA_j) & = \textstyle{ \sum_{l} G_{il}(\vA_{jl}) } \\
      & = \textstyle{ \sum_{l} \min_{r} \{g_{jl} d_{jr} / d_{ir} \} } \\
      & \le \textstyle{ \sum_{l} g_{jl} } = G_i(\vA_i) ~,
  \end{align*}
  where the inequality holds because
  \begin{equation*}
	  \min_r \{ d_{jr} / d_{ir} \} \le d_{jr_i^*} / d_{i r_i^*} \le 1~,
  \end{equation*}
  where $r_i^*$ is user $i$'s global dominant resource.
\qed

We next show that DRFH leads to an efficient allocation under which no user can
improve its allocation without decreasing that of the others.

\begin{proposition}[Pareto optimality]
  \label{prop:po}
  The DRFH allocation obtained by solving (\ref{eq:h-drf}) is Pareto optimal.
\end{proposition}

{\bf Proof:}
  Let $\{ g_{il} \}$, and the corresponding $g$, be the solution to problem (\ref{eq:h-drf}). For all
  user $i$, its DRFH allocation in server $l$ is $\vA_{il} = g_{il} \vd_i$. Since
  (\ref{eq:maxmin-fair}) and (\ref{eq:h-drf}) are equivalent, $\{ g_{il} \}$
  also solves (\ref{eq:maxmin-fair}), with $g$ being the maximum value of the
  objective of \eqref{eq:maxmin-fair}.
  
  Assume, by way of contradiction, that allocation $\vA$ is not Pareto optimal,
  {\em i.e.,} there exists some allocation $\vA'$, such that $N_i(\vA'_i) \ge
  N_i(\vA_i)$ for all user $i$, and for some user $j$ we have strict
  inequality: $N_j(\vA'_j) > N_j(\vA_j)$. Equivalently, this implies that
  $G_i(\vA'_i) \ge G_i(\vA_i)$ for all user $i$, and $G_j(\vA'_j) > G_j(\vA_j)$
  for user $j$.  Without loss of generality, let $\vA'$ be
  non-wasteful. By Lemma~\ref{lem:nonwaste}, for all user $i$ and server $l$,
  there exists some $g'_{il}$ such that $\vA'_{il} = g'_{il} \vd_i$.  We show
  that based on $\{ g'_{il} \}$, one can construct some $\{ \hat{g}_{il} \}$
  such that $\{ \hat{g}_{il} \}$ is a feasible solution to
  (\ref{eq:maxmin-fair}), yet leads to a higher objective than $g$, 
  contradicting the fact that $\{ g_{il} \}$ optimally solve
  (\ref{eq:maxmin-fair}).

  To see this, consider user $j$. We have
  \begin{equation*}
    \textstyle{G_j(\vA_j) = \sum_l g_{jl} = g < G_j(\vA'_j) = \sum_l g'_{jl}}.
  \end{equation*}
  For user $j$, there exists a server $l_0$ and some $\epsilon > 0$, such that
  after reducing $g'_{jl_0}$ to $g'_{jl_0} -
  \epsilon$, the resulting global dominant share remains higher than $g$, {\em i.e.,} $\sum_l g'_{jl} -
  \epsilon \ge g$. This leads to at least $\epsilon \vd_j$ idle resources in
  server $l_0$. We construct $\{ \hat{g}_{il} \}$ by redistributing these idle
  resources to all users.
  
  Denote by $\{ g''_{il} \}$ the dominant share after reducing
  $g'_{jl_0}$ to $g'_{jl_0} - \epsilon$, {\em i.e.,}
  \begin{equation*}
    g''_{il} = \left\{
    \begin{array}{ll}
      g'_{jl_0} - \epsilon, & i = j, l = l_0; \\
      g'_{il}, & o.w.
    \end{array}
    \right.
  \end{equation*}
  The corresponding non-wasteful allocation is $\vA''_{il} =
  g''_{il} \vd_i$ for all user $i$ and server $l$. Note that allocation $\vA''$ is preferred over
  the original allocation $\vA$ by all users, {\em i.e.,} for all user $i$, we have
  \begin{equation*}
    G_i(\vA''_i) = \sum_l g''_{il} = \left\{
    \begin{array}{ll}
      \sum_l g'_{jl} - \epsilon \ge g = G_j(\vA_j), & i = j; \\
      \sum_l g'_{il} = G_i(\vA'_i) \ge G_i(\vA_i), & o.w.
    \end{array}
    \right.
    \nonumber
  \end{equation*}

  We now construct $\{ \hat{g}_{il} \}$ by redistributing the $\epsilon \vd_j$
  idle resources in server $l_0$ to all users, each increasing its global
  dominant share $g''_{il_0}$ by $\delta = \min_r \{ \epsilon d_{jr} /
	  \sum_i d_{ir}\}$, {\em i.e.},
  \begin{equation*}
    \hat{g}_{il} = \left\{
      \begin{array}{ll}
	g''_{il_0} + \delta, & l = l_0; \\
	g''_{il}, & o.w.
      \end{array}
    \right.
  \end{equation*}
  It is easy to check that $\{ \hat{g}_{il} \}$ remains a feasible allocation.
  To see this, it suffices to check server $l_0$. For all its resource
  $r$, we have
  \begin{equation*}
    \begin{split}
      \textstyle{ \sum_i \hat{g}_{il_0} d_{ir} } & 
      = \textstyle{ \sum_i (g''_{il_0} + \delta) d_{ir} } \\
      & = \textstyle{ \sum_i g'_{il_0} d_{ir} - \epsilon d_{jr} + \delta \sum_i d_{ir} } \\
      & \le \textstyle{ c_{l_0 r} - ( \epsilon d_{jr} - \delta \sum_i d_{ir} ) } \le c_{l_0 r}~.
    \end{split}
    \nonumber
  \end{equation*}
  where the first inequality holds because $\vA'$ is a feasible allocation. 

  On the other hand, for all user $i \in U$, we have
  \begin{equation*}
    \begin{split}
      \textstyle{ \sum_l \hat{g}_{il} } & 
      = \textstyle{ \sum_l g''_{il} + \delta } \\
      & = G_i(\vA''_i) + \delta \\
      & \ge G_i(\vA_i) + \delta > g~.
    \end{split}
    \label{eq:max-exceeded}
  \end{equation*}
  This contradicts the premise that $g$ is optimal for (\ref{eq:maxmin-fair}).
\qed

For now, all our discussions are based on a critical assumption that all users
truthfully report their resource demands. However, in a real-world system, it is
common to observe users to attempt to manipulate the scheduler by misreporting
their resource demands, so as to receive more allocation \cite{Ghodsi11a,
	Ghodsi13a}.  More often than not, these strategic behaviours would
significantly hurt those honest users and reduce the number of their tasks
scheduled, inevitably leading to a fairly inefficient allocation outcome.
Fortunately, we show by the following proposition that DRFH is immune to these
strategic behaviours, as reporting the true demand is always the best strategy
for every user, irrespective of the others' behaviour.

\begin{proposition}[Truthfulness]
	\label{prop:truthful}
  The DRFH allocation obtained by solving (\ref{eq:h-drf}) is truthful.
\end{proposition}

{\bf Proof:}
  For any user $i$, fixing all other users' clamed demands $\vd'_{-i} = (\vd'_{1}, \dots,
  \vd'_{i-1}, \vd'_{i+1}, \dots, \vd'_n)$ (which may not be their true demands),
  let $\vA$ be the resulting allocation when $i$ truthfully reports its demand
  $\vd_i$, that is, $\vA_{il} = g_{il} \vd_i$ and $\vA_{jl} = g_{jl} \vd'_j$
  for all user $j \neq i$ and server $l$, where $g_{il}$ and $g_{jl}$ are the global dominant shares
  users $i$ and $j$ receive on server $l$ under $\vA_{il}$ and $\vA_{jl}$, respectively. 
  Similarly, let $\vA'$ be
  the resulting allocation when user $i$ misreports its demand as $\vd'_i$.
  Let $g$ and $g'$ be the global dominant share user $i$
  receives under $\vA_i$ and $\vA'_i$, respectively. We check the following two
  cases and show that $G_i(\vA'_i) \le G_i(\vA_i)$, which is equivalent to
  $N_i(\vA'_i) \le N_i(\vA_i)$.

  {\em Case 1:} $g' \le g$. In this case, let $\rho_i = \min_r \{ d'_{ir} /
  d_{ir} \}$ be defined for user $i$. Clearly,
  \begin{equation*}
    \textstyle{ \rho_i = \min_r \{ d'_{ir} / d_{ir} \} \le d'_{i r_i^*} / d_{i
    r_i^*} \le 1 }~,
  \end{equation*}
  where $r_i^*$ is the dominant resource of user $i$. We then have
  \begin{align*}
    G_i(\vA'_i) & = \textstyle{ \sum_l G_{il}(\vA'_{il}) } \\
    & = \textstyle{ \min_r \{ d'_{ir} / d_{ir} \} \sum_l g'_{il}} \\
    & = \textstyle{ \rho_i g' } \le g = G_i(\vA_i) ~.
  \end{align*}

  {\em Case 2:} $g' > g$. 
  For all user $j \neq i$, when user $i$ truthfully reports its demand, let $G_j(\vA_j, \vd'_j)$
  be the global dominant share of user $j$ w.r.t. its {\em claimed demand} $\vd'_j$, {\em i.e.,}
  \begin{equation*}
    \textstyle{ G_j(\vA_j, \vd'_j) = \sum_l \min_r \{ g_{jl} d'_{jr} / d'_{jr} \} 
    = \sum_l g_{jl} = g } ~.
    \nonumber
  \end{equation*}
  Similarly, when user $i$ misreports, let $G_j(\vA'_j, \vd'_j)$
  be the global dominant share of user $j$ w.r.t. its claimed demand $\vd'_j$, {\em i.e.,}
  \begin{equation*}
    \textstyle{ G_j(\vA'_j, \vd'_j) = \sum_l \min_r \{ g'_{jl} d'_{jr} / d'_{jr} \} 
    = \sum_l g'_{jl} = g' }~,
    \nonumber
  \end{equation*}
  As a result,
  \begin{equation*}
    G_j(\vA'_j, \vd'_j) > G_j(\vA_j, \vd'_j), \quad \forall j \neq i ~.
  \end{equation*}
  We must have
  \begin{equation*}
    G_i(\vA'_i) < G_i(\vA_i)~.
  \end{equation*}
  Otherwise, allocation $\vA'$ is preferred over $\vA$ by all users and is
  strictly preferred by user $j \neq i$ w.r.t.~the claimed demands $(\vd'_{-i},
  \vd_i)$. This contradicts the Pareto optimality of DRFH allocation. (Recall that allocation
  $\vA$ is an DRFH allocation given the claimed demands $(\vd'_{-i}, \vd_i)$. )
\qed

\subsection{Analysis of Important Properties}

In addition to the three essential properties shown in the previous subsection,
DRFH also provides a number of other important properties.  First, since DRFH
generalizes DRF to heterogeneous environments, it naturally reduces to the DRF
allocation when there is only one server contained in the system,
where the global dominant resource defined in DRFH is exactly the same as the
dominant resource defined in DRF.

\begin{proposition}[Single-server DRF]
  The DRFH leads to the same allocation as DRF when all resources are
  concentrated in one server.
\end{proposition}

Next, by definition, we see that both single-resource fairness and bottleneck
fairness trivially hold for the DRFH allocation. We hence omit the proofs of
the following two propositions.

\begin{proposition}[Single-resource fairness]
  The DRFH allocation satisfies single-resource fairness.
\end{proposition}

\begin{proposition}[Bottleneck fairness]
  The DRFH allocation satisfies bottleneck fairness.
\end{proposition}

Finally, we see that when a user leaves the system and relinquishes all its
allocations, the remaining users will not see any reduction of the number of
tasks scheduled. Formally,

\begin{proposition}[Population monotonicity]
  The DRFH allocation satisfies population monotonicity.
\end{proposition}

{\bf Proof:}
  Let $\vA$ be the resulting DRFH allocation, then for all user $i$ and server $l$, $\vA_{il} = g_{il} \vd_i$ and
  $G_i(\vA_i) = g$, where $\{ g_{il} \}$ and $g$ solve
  (\ref{eq:h-drf}). Suppose user $j$ leaves the system, changing the
  resulting DRFH allocation to $\vA'$. By DRFH, for all user $i \neq j$ and server $l$, we have $\vA'_{il}
  = g'_{il} \vd_i$ and $G_i(\vA'_i) = g'$, where $\{
	  g'_{il} \}_{i \neq j}$ and $g'$ solve the following optimization problem:
  \begin{equation}
    \label{eq:h-drf-less-user}
    \begin{split}
	    \textstyle{ \max_{g'_{il}, i \neq j} } & \quad g' \\
	    \mbox{s.t.} & \quad \textstyle{ \sum_{i \neq j} g'_{il} d_{ir} \le c_{lr}, \forall l \in S, r \in R~, } \\
	    & \quad \textstyle{ \sum_{l \in U} g'_{il} = g', \forall i \neq j ~. }
    \end{split}
  \end{equation}

  To show $N_i(\vA'_i) \ge N_i(\vA_i)$ for all user $i \neq j$, it is equivalent to 
  prove $G_i(\vA'_i) \ge G_i(\vA_i)$. It is easy to verify that $g, \{ g_{il} \}_{i
	  \neq j}$ satisfy all the constraints of (\ref{eq:h-drf-less-user})
  and are hence feasible to \eqref{eq:h-drf-less-user}. As a result, $g' \ge
  g$. This is exactly $G_i(\vA'_i) \ge G_i(\vA_i)$.
\qed

\subsection{Discussions of Sharing Incentive}
\label{sec:sharing-incentive}

In addition to the aforementioned properties, sharing incentive is another
important allocation property that has been frequently mentioned in the
literature, {\em e.g.,} \cite{Ghodsi11a, Joe-Wong12a, Dolev12a, Parkes12a,
	Ghodsi13a}. It ensures that every user's allocation is at least as good
as that obtained by evenly partitioning the entire resource pool. When the
system contains only a single server, this property is well defined, as evenly
dividing the server's resources leads to a {\em unique} allocation. However,
for the system containing multiple heterogeneous servers, there is an {\em
	infinite} number of ways to evenly divide the resource pool, and it is
unclear which one should be chosen as the benchmark for comparison.  For
example, in Fig.~\ref{fig:drf-not-po}, two users share the system with 14 CPUs
and 14 GB memory in total. The following two allocations both allocate each
user 7 CPUs and 7 GB memory: (a) User 1 is allocated 1/2 resources of server 1
and 1/2 resources of server 2, while user 2 receives the rest; (b) user 1 is
allocated (1.5 CPUs, 5.5 GB) in server 1 and (5.5 CPUs, 1.5 GB) in server 2,
while user 2 receives the rest. 

One might think that allocation (a) is a more reasonable benchmark as it allows
all $n$ users to have an equal share of every server, each receiving $1/n$ of
the server's resources.  However, this benchmark has little practical meaning:
With a large $n$, each user will only receive a small fraction of resources on
each server, which likely cannot be utilized by any computing task. In other
words, having a small slice of resources in each server is essentially
meaningless. We therefore consider another benchmark that is more practical. 

Since cloud systems are constructed by pooling hundreds of thousands of servers
\cite{Armbrust10a, Reiss12a}, the number of users is typically far smaller than
the number of servers \cite{Ghodsi11a, Ghodsi13a}, {\em i.e.}, $k \gg n$. An
equal division would allocate to each user $k/n$ servers drawn from the same
distribution of the system's server configurations. For each user, the
allocated $k/n$ servers are then treated as a {\em dedicated cloud} that is
exclusive to the user. The number of tasks scheduled on this dedicated cloud is
then used as a benchmark and is compared to the number of tasks scheduled in
the original cloud computing system shared with all other users. We will evaluate
such a sharing incentive property via trace-driven simulations in
Sec.~\ref{sec:sim}.

\section{Practical Considerations}
\label{sec:prac}

So far, all our discussions are based on several assumptions that may not be
the case in a real-world system. In this section, we relax these assumptions and
discuss how DRFH can be implemented in practice.

\subsection{Weighted Users with a Finite Number of Tasks}

In the previous sections, users are assumed to be assigned equal weights and
have infinite computing demands. Both assumptions can be easily removed with
some minor modifications of DRFH. 

When users are assigned uneven weights, let $w_i$ be the weight associated with
user $i$. DRFH seeks an allocation that achieves the {\em weighted max-min
	fairness} across users. Specifically, we maximize the minimum {\em
	normalized global dominant share} (w.r.t the weight) of all users
under the same resource constraints as in
\eqref{eq:max-min}, {\em i.e.,}
\begin{equation*}
  \begin{split}
	  \max_{\vA} & \quad \min_{i \in U} G_i(\vA_i) / w_i \\
	  \mbox{s.t.} & \quad \sum_{i \in U} A_{ilr} \le c_{lr}, \forall l \in S, r \in R~.
  \end{split}
\end{equation*}

When users have a finite number of tasks, the DRFH allocation is computed
iteratively. In each round, DRFH increases the global dominant share allocated
to all {\em active users}, until one of them has all its tasks scheduled, after
which the user becomes inactive and will no longer be considered in the
following allocation rounds. DRFH then starts a new iteration and repeats the
allocation process above, until no user is active or no more resources could be
allocated to users. Our analysis presented in Sec.~\ref{sec:drfc} also extends
to weighted users with a finite number of tasks.

\subsection{Scheduling Tasks as Entities}
\label{sec:task-scheduling}

Until now, we have assumed that all tasks are divisible. In a real-world system,
however, fractional tasks may not be accepted. To schedule tasks as entities,
one can apply {\em progressive filling} as a simple implementation of DRFH.
That is, whenever there is a scheduling opportunity, the scheduler always
accommodates the user with the lowest global dominant share. To do this, it
picks the first server that fits the user's task. While this First-Fit
algorithm offers a fairly good approximation to DRFH, we propose another simple heuristic that can lead
to a better allocation with higher resource utilization.

Similar to First-Fit, the heuristic also chooses user $i$ with
the lowest global dominant share to serve. However, instead of randomly picking
a server, the heuristic chooses the ``best'' one that most suitably matches user $i$'s tasks, and is
hence referred to as the {\em Best-Fit} DRFH.
Specifically, for user $i$ with resource demand vector $\vD_i = (D_{i1},
\dots, D_{im})^T$ and a server $l$ with {\em available resource vector}
$\bar{\vc}_l = (\bar{c}_{l1}, \dots, \bar{c}_{lm})^T$, where $\bar{c}_{lr}$ is
the share of resource $r$ remaining available in server $l$, we define the following
{\em heuristic function} to measure the task's fitness for the server:
\begin{equation}
	H(i, l) = \norm{ \vD_i / D_{i1} - \bar{\vc}_l / \bar{c}_{l1}}_1~,
\end{equation}
where $\norm{ \cdot }_1$ is the $L^1$-norm. Intuitively, the smaller $H(i, l)$,
the more similar the resource demand vector $\vD_i$ appears to the server's
available resource vector $\bar{\vc}_l$, and the better fit user $i$'s task is
for server $l$.  For example, a CPU-heavy task is more suitable to run in a
server with more available CPU resources. Best-Fit DRFH schedules user $i$'s
tasks to server $l$ with the least $H(i,l)$.
We evaluate both First-Fit DRFH and Best-Fit DRFH via trace-driven
simulations in the next section.

\section{Simulation Results}
\label{sec:sim}

In this section, we evaluate the performance of DRFH via extensive simulations
driven by Google cluster-usage traces \cite{GoogleTrace}. The traces contain
resource demand/usage information of over 900 users ({\em i.e.,} Google
services and engineers) on a cluster of 12K servers. The server configurations
are summarized in Table~\ref{tbl:server-config}, where the CPUs and memory of
each server are normalized so that the maximum server is 1. Each user submits
computing jobs, divided into a number of tasks, each requiring a set of
resources ({\em i.e.,} CPU and memory). From the traces, we extract the
computing demand information --- the required amount of resources and task
running time --- and use it as the demand input of the allocation algorithms
for evaluation.

\begin{figure}[tb]
  \centering
  \includegraphics[width=0.47\textwidth]{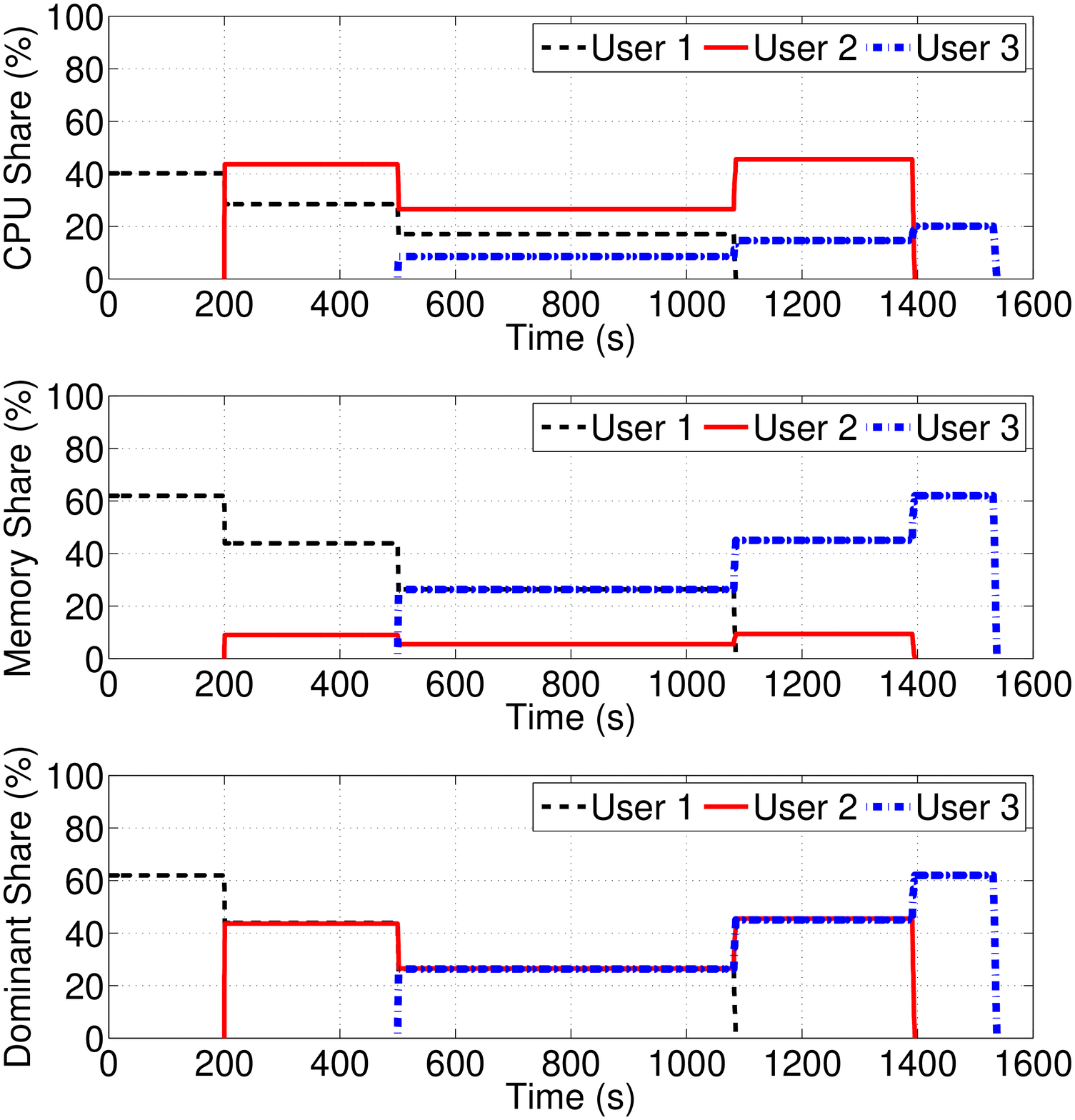}
  \vspace{-.1in}
  \caption{CPU, memory, and global dominant share for three users on a 100-server system with 52.75 CPU
	  units and 51.32 memory units in total.
  }
  \label{fig:dynamic-alloc}
  \vspace{-.1in}
\end{figure}

{\bf Dynamic allocation:} Our first evaluation focuses on the allocation
fairness of the proposed Best-Fit DRFH when users dynamically join and depart
the system. We simulate 3 users submitting tasks with different resource
requirements to a small cluster of 100 servers. The server configurations are randomly
drawn from the distribution of Google cluster servers in
Table~\ref{tbl:server-config}, leading to a resource pool containing 52.75 CPU units and
51.32 memory units in total. User 1 joins the system at the beginning, requiring 0.2
CPU and 0.3 memory for each of its task. As shown in
Fig.~\ref{fig:dynamic-alloc}, since only user 1 is active at the beginning, it
is allocated 40\% CPU share and 62\% memory share. This allocation continues
until 200~s, at which time user 2 joins and submits CPU-heavy tasks, each
requiring 0.5 CPU and 0.1 memory. Both users now compete for computing resources,
leading to a DRFH allocation in which both users receive 44\% global
dominant share. At 500~s, user 3 starts to submit memory-intensive tasks, each
requiring 0.1 CPU and 0.3 memory. The algorithm now allocates the same global
dominant share of 26\% to all three users until user 1 finishes its tasks and
departs at 1080~s. After that, only users 2 and 3 share the system, each
receiving the same share on their global dominant resources.  A similar process
repeats until all users finish their tasks. Throughout the simulation, we see
that the Best-Fit DRFH algorithm precisely achieves the DRFH allocation at all times.

\begin{table}[t]
   \centering
   \renewcommand{\arraystretch}{0.99}
   \footnotesize
   \caption{Resource utilization of the Slots scheduler with different slot sizes.}
   \vspace{-2mm}
   \begin{tabular}{|c||c|c|}
     \hline
     {\bf Number of Slots} & {\bf CPU Utilization} & {\bf Memory Utilization} \\
     \hline
     10 per maximum server & 35.1\% & 23.4\% \\
     \hline
     12 per maximum server & 42.2\% & 27.4\% \\
     \hline
     14 per maximum server & 43.9\% & 28.0\% \\
     \hline
     16 per maximum server & 45.4\% & 24.2\% \\
     \hline
     20 per maximum server & 40.6\% & 20.0\% \\
     \hline
   \end{tabular}
   \label{tbl:slot-util}
   \vspace{-.1in}
\end{table}

{\bf Resource utilization:} We next evaluate the resource utilization of the
proposed Best-Fit DRFH algorithm. We take the 24-hour computing demand data from the
Google traces and simulate it on a smaller cloud computing system of 2,000 servers so that
fairness becomes relevant. The server configurations are randomly drawn from the
distribution of Google cluster servers in Table~\ref{tbl:server-config}. We
compare Best-Fit DRFH with two other benchmarks, the traditional Slots
schedulers that schedules tasks onto slots of servers ({\em e.g.,} Hadoop Fair
Scheduler \cite{HadoopFS}), and the
First-Fit DRFH that chooses the first server that fits the task. For the
former, we try different slot sizes and chooses the one with the highest CPU
and memory utilization.  Table~\ref{tbl:slot-util} summarizes our observations,
where dividing the maximum server (1 CPU and 1 memory in
Table~\ref{tbl:server-config}) into 14 slots leads to the highest overall
utilization.

\begin{figure}[tb]
  \centering
  \includegraphics[width=0.47\textwidth]{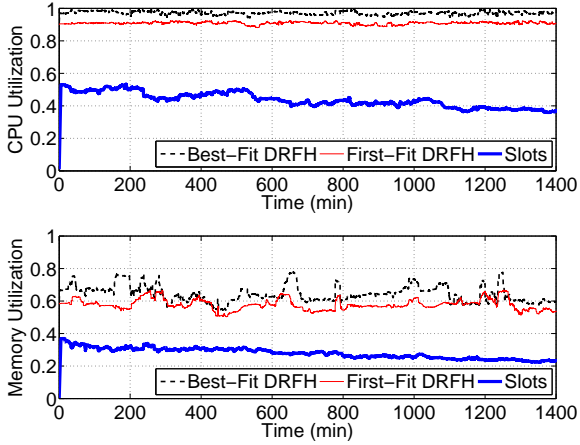}
  \vspace{-.1in}
  \caption{Time series of CPU and memory utilization.
  }
  \label{fig:res-util}
  \vspace{-.1in}
\end{figure}

Fig.~\ref{fig:res-util} depicts the time series of CPU and memory utilization
of the three algorithms. We see that the two DRFH implementations significantly
outperform the traditional Slots scheduler with much higher resource
utilization, mainly because the latter ignores the heterogeneity of both
servers and workload. This observation is consistent with findings in the
homogeneous environment where all servers are of the same hardware
configurations \cite{Ghodsi11a}. As for the DRFH implementations, we see that
Best-Fit DRFH leads to uniformly higher resource utilization than the First-Fit
alternative at all times.

\begin{figure}[tb]
  \centering
  \subfloat[CDF of job completion times.]
  {
    \label{fig:comp-time-cdf}
    \includegraphics[width=0.23\textwidth]{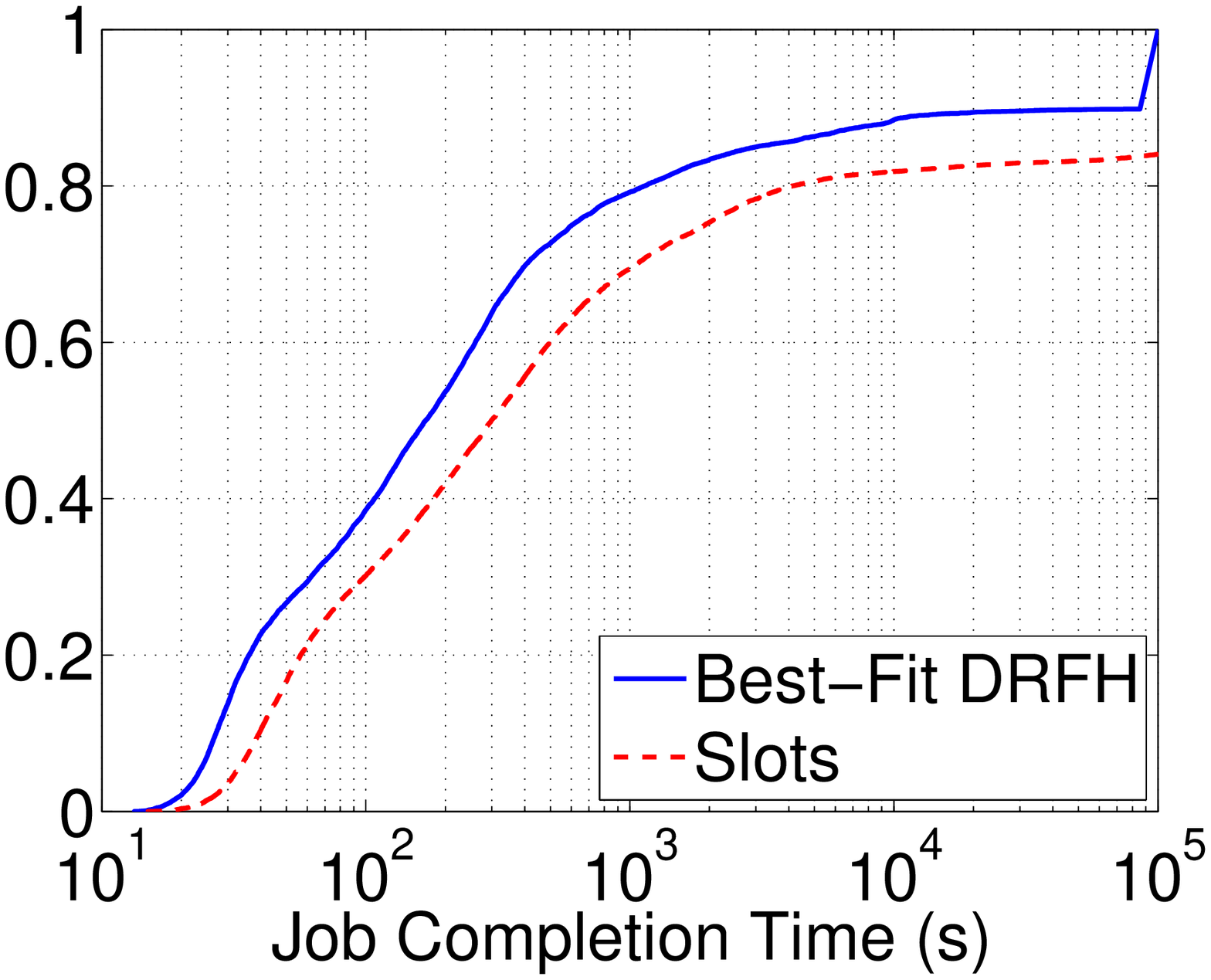}
  }
  \subfloat[Job completion time reduction.]
  {
    \label{fig:comp-time-red}
    \includegraphics[width=0.23\textwidth]{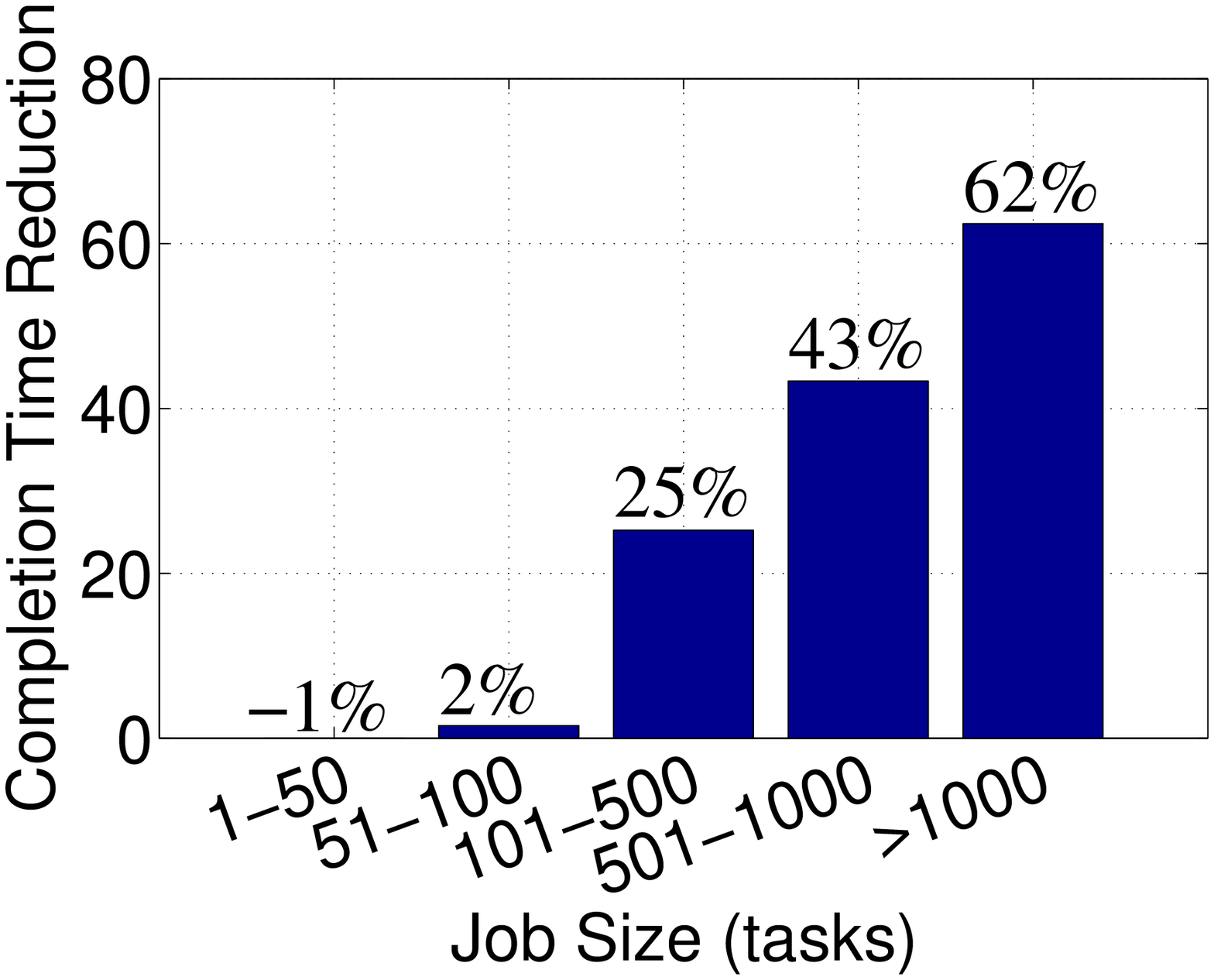}
  }
  \caption{DRFH improvements on job completion times over Slots scheduler.
  }
  \label{fig:comp-time}
  \vspace{-.1in}
\end{figure}

The high resource utilization of Best-Fit DRFH naturally translates to shorter
job completion times shown in Fig.~\ref{fig:comp-time-cdf}, where the CDFs of
job completion times for both Best-Fit DRFH and Slots scheduler are depicted.
Fig.~\ref{fig:comp-time-red} offers a more detailed breakdown, where jobs are
classified into 5 categories based on the number of its computing tasks, and
for each category, the mean completion time reduction is computed. While DRFH
shows no improvement over Slots scheduler for small jobs, a significant
completion time reduction has been observed for those containing more tasks.
Generally, the larger the job is, the more improvement one may expect. Similar
observations have also been found in the homogeneous environments
\cite{Ghodsi11a}.

\begin{figure}[tb]
       	\centering
	\begin{minipage}[t]{.47\linewidth}
	       	\centering
	       	\includegraphics[width=\linewidth]{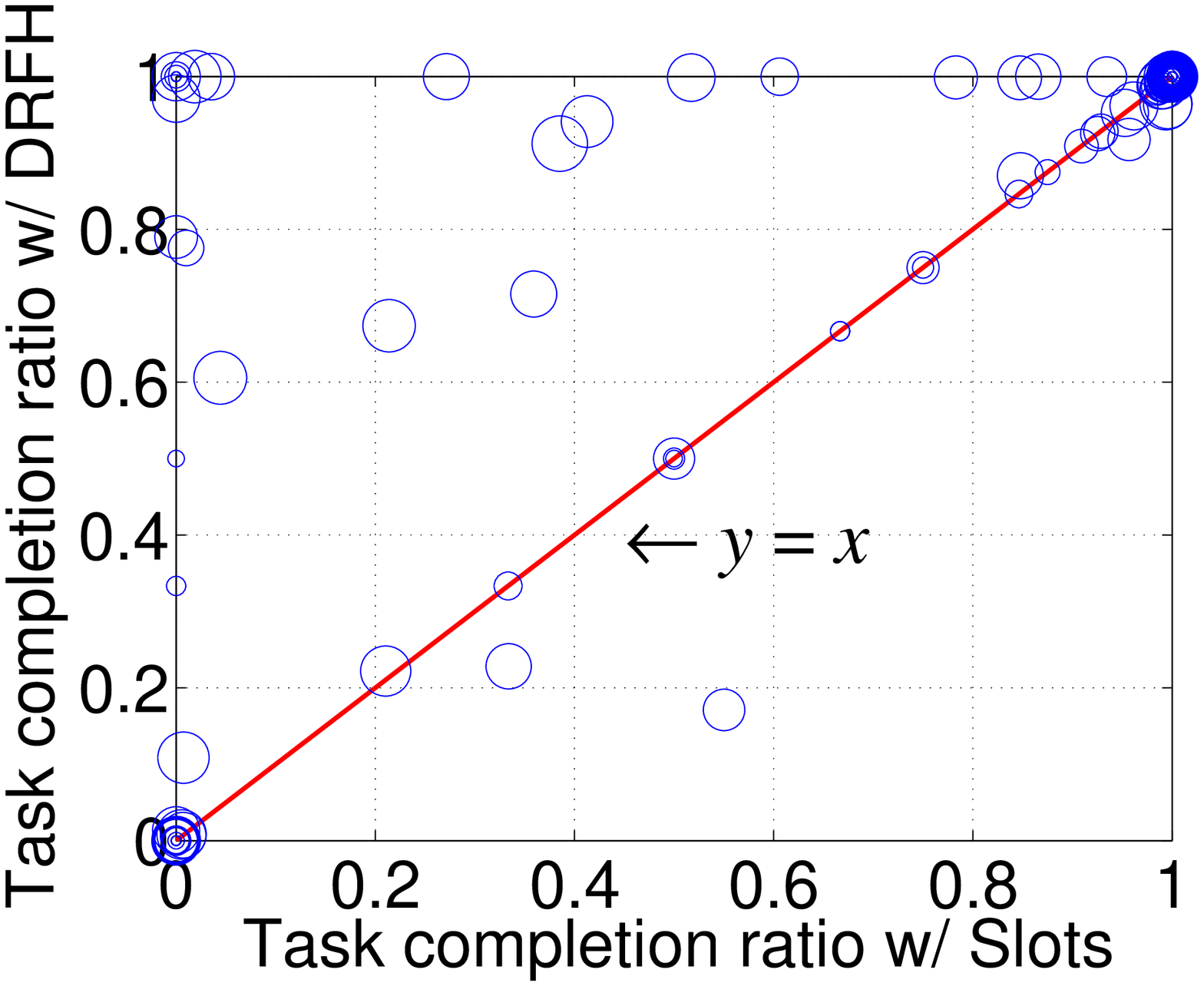}
		\caption{Task completion ratio of users using Best-Fit DRFH and Slots schedulers,
			respectively. Each bubble's size is logarithmic to the number of
			tasks the user submitted.}
		\label{fig:tasks-completed}
       	\end{minipage}%
	\hfill
	\begin{minipage}[t]{.47\linewidth}
	       	\centering
	       	\includegraphics[width=\linewidth]{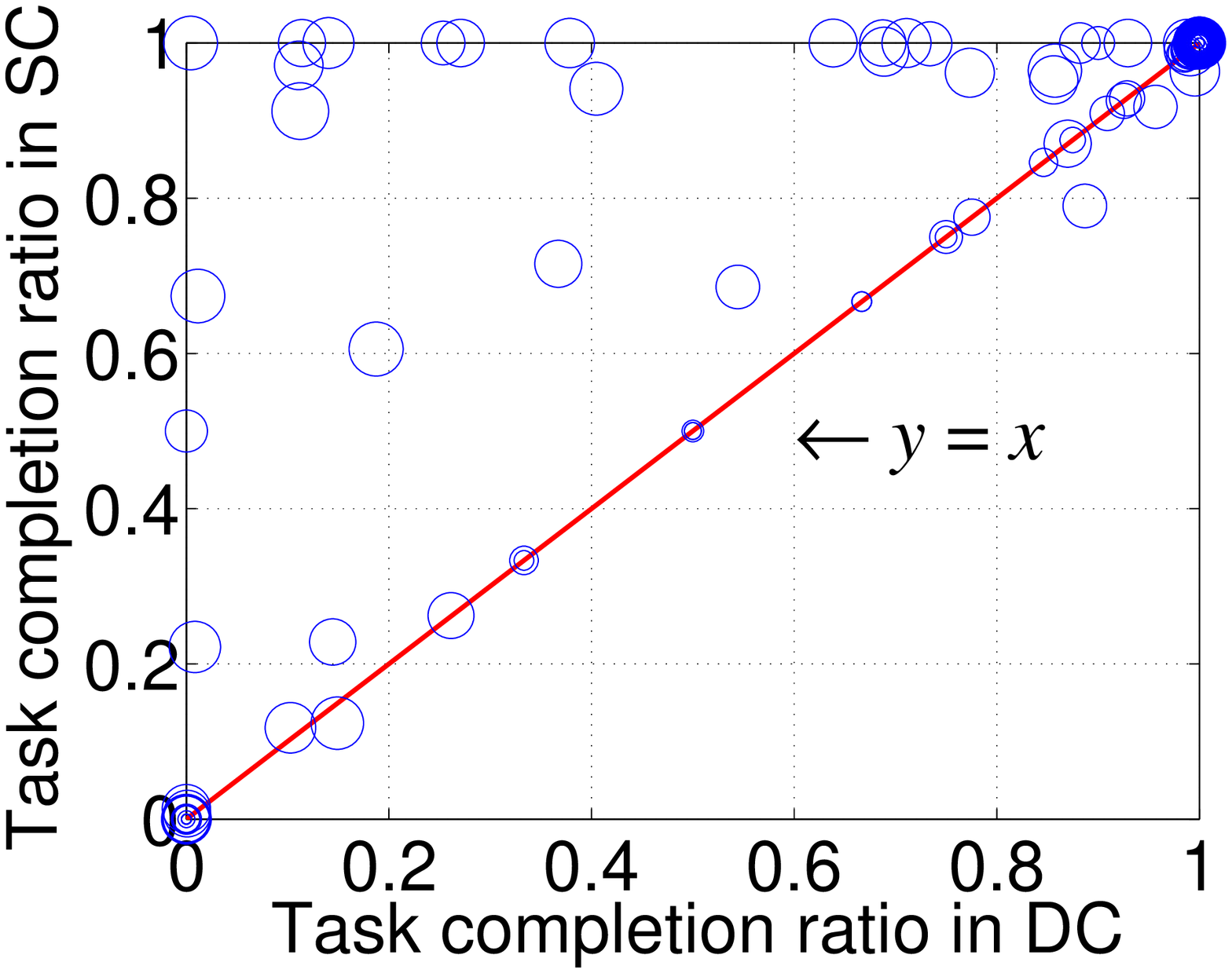}
		\caption{Task completion ratio of users running on dedicated
			clouds (DCs) and the shared cloud (SC). Each circle's radius is
			logarithmic to the number of tasks submitted.}
		\label{fig:sharing-incent}
       	\end{minipage}
	\vspace{-.1in}
\end{figure}

Fig.~\ref{fig:comp-time} does not account for partially completed jobs and
focuses only on those having all tasks finished in both Best-Fit and Slots. As
a complementary study, Fig.~\ref{fig:tasks-completed} computes the task
completion ratio --- the number of tasks completed over the number of tasks
submitted --- for every user using Best-Fit DRFH and Slots schedulers,
respectively. The radius of the circle is scaled logarithmically to the number
of tasks the user submitted.  We see that Best-Fit DRFH leads to higher task
completion ratio for almost all users. Around 20\% users have all their tasks
completed under Best-Fit DRFH but do not under Slots.

{\bf Sharing incentive:}
Our final evaluation is on the sharing incentive property of DRFH. As mentioned
in Sec.~\ref{sec:sharing-incentive}, for each user, we run its computing tasks
on a dedicated cloud (DC) that is a proportional subset of the original shared
cloud (SC). We then compare the task completion ratio in DC with that obtained in
SC. Fig.~\ref{fig:sharing-incent} illustrates the results.
While DRFH does not guarantee 100\% sharing incentive for all users, it
benefits most of them by pooling their DCs together. In
particular, only 2\% users see fewer tasks finished in the shared environment.
Even for these users, the task completion ratio decreases only slightly, as can be seen from Fig.~\ref{fig:sharing-incent}.

\section{Concluding Remarks}
\label{sec:conclusion}

In this paper, we study a multi-resource allocation problem in a heterogeneous
cloud computing system where the resource pool is composed of a large number of
servers with different configurations in terms of resources such as processing, memory, and
storage. The proposed multi-resource allocation mechanism, known as DRFH,
equalizes the global dominant share allocated to each user, and hence
generalizes the DRF allocation from a single server to multiple heterogeneous
servers. We analyze DRFH and show that it retains almost all desirable
properties that DRF provides in the single-server scenario. Notably, DRFH is
envy-free, Pareto optimal, and truthful. We design a Best-Fit heuristic that
implements DRFH in a real-world system. Our large-scale simulations driven by
Google cluster traces show that, compared to the traditional single-resource
abstraction such as a slot scheduler, DRFH achieves significant improvements
in resource utilization, leading to much shorter job completion
times.

\bibliographystyle{IEEEtran}
\bibliography{main}

\end{document}